\documentclass[10pt,letterpaper]{article}
\usepackage[top=0.85in,left=1.75in,footskip=0.75in]{geometry}

\usepackage{amsmath,amssymb}

\usepackage{changepage}

\usepackage{textcomp,marvosym}

\usepackage{cite}

\usepackage{nameref,hyperref}

\usepackage[nopatch=eqnum]{microtype}
\DisableLigatures[f]{encoding = *, family = * }

\usepackage[table]{xcolor}

\usepackage{array}

\newcolumntype{+}{!{\vrule width 2pt}}

\newlength\savedwidth

\raggedright
\usepackage[aboveskip=1pt,labelfont=bf,labelsep=period,justification=raggedright,singlelinecheck=off]{caption}

\makeatletter
\renewcommand{\@biblabel}[1]{\quad#1.}
\makeatother

\usepackage{lastpage,fancyhdr,graphicx}
\usepackage{epstopdf}
\fancyheadoffset[L]{1.25in}
\fancyfootoffset[L]{1.25in}
\begin{document}
	\vspace*{0.2in}
	
	\begin{flushleft}
		{\Large
			\textbf\newline{Applications of the Vendi score in genomic epidemiology}
		}
		\newline
		\\
		Bjarke Frost Nielsen\textsuperscript{1,2*},
		Amey P. Pasarkar\textsuperscript{3,6},
		Qiqi Yang\textsuperscript{4},
		Bryan T. Grenfell\textsuperscript{4},
		{Adji Bousso Dieng}\textsuperscript{5,6*}
		\\
		\bigskip
		\textbf{1} High Meadows Environmental Institute, Princeton University, Princeton, NJ, USA
		\\
		\textbf{2} Niels Bohr Institute, University of Copenhagen, Copenhagen, Denmark
		\\
		\textbf{3} Lewis-Sigler Institute For Integrative Genomics, Princeton University
		\\
		\textbf{4} Department of Ecology and Evolutionary Biology, Princeton University, Princeton, NJ, USA
		\\
		\textbf{5} Department of Computer Science, Princeton University, Princeton, NJ, USA\\
		\textbf{6} Vertaix, Princeton, NJ, USA
		\bigskip

		* bjarke@princeton.edu, adji@princeton.edu
		
	\end{flushleft}
	\section*{Abstract}
	The Vendi score (VS), a diversity metric recently conceived in the context of machine learning, with applications in a wide range of fields, has a few distinct advantages over the metrics commonly used in ecology. It is classification-\hspace{0pt}independent, incorporates abundance information, and has a tunable sensitivity to rare/\hspace{0pt}abundant types. 
	Using rich COVID-19 sequence data as a paradigm, we develop methods for applying the VS to time-resolved sequence data.  We show how the VS allows for characterization of the overall diversity of circulating viruses and for discernment of emerging variants prior to formal identification. Furthermore, applying the VS to phylogenetic trees provides a convenient overview of within-clade diversity which can aid viral variant detection.
	
	\section*{Author summary}
	We present techniques to apply the Vendi score, a recently developed diversity measure, to viral genomic epidemiology. The Vendi score is highly flexible and unsupervised, meaning that it does not rely on predefined categories such as lineages or variants. This allows us to detect subtle shifts in viral diversity, including the early signs of emerging variants. The Vendi score is efficient and straight-forward to apply: it requires only the raw sequence data and a chosen similarity function. By analyzing SARS-CoV-2 genomes, we show how the Vendi score can highlight low-diversity clusters of viral sequences -- potentially signaling emerging variants before they are formally recognized.
	
	
	\section*{Introduction}
	Diversity measures in ecology tend to rely on a pre-existing classification, into e.g. species or variants. With rapidly evolving pathogens such as RNA viruses, as well as dramatically increased pathogen sequencing efforts, there is a pressing need
	for flexible and informative diversity measures that can be applied in real time as samples become available. When rapid response is essential, as in outbreak control, tools which bypass potentially laborious classification processes have the potential to strengthen surveillance. Historically, species classification of viruses has been contentious, and this difficulty continues at lower taxonomic levels.  Over the years, the International Committee on Taxonomy of Viruses (ICTV) has laid out a succession of definitions of viral \textit{species} \cite{fauquet1999taxonomy,vanRegenmortel1992}, since viruses do not fit neatly into traditional species concepts such the Mayr definition \cite{mayr1970populations}, which focuses on sexually reproducing populations. The current definition states that \textit{``[a] species is a monophyletic group of MGEs [Mobile Genetic Elements] whose properties can be distinguished from those of other species by multiple criteria.''}\cite{walker2021changes} 
	Below the species level, similar challenges of demarcation arise. Indeed, no universal classification approach exists \cite{rambaut2020dynamic}, and monophyletic groups may be referred to as (sub)types, genotypes, variants, sub-variants etc. While classification of viral variants is of course indispensable and is largely what allows tracking the phenotypic changes in a pathogen over time, there is a need for tools which allow characterization of changes in viral populations before classification is finalized.
	
	The Vendi score \cite{friedman2022vendi} is a flexible and tunable diversity score that requires no pre-classification, and instead depends only on a relevant similarity metric being defined.
	The high generality of the Vendi score -- owing to relying only on a notion of similarity -- has led to application to a diverse set of problems ranging from molecular simulation \cite{pasarkar2023molecular}, evaluating and improving generative machine learning models \cite{friedman2022vendi,rezaei2025vendi,pasarkar2023cousins,rezaei2025alpha}, experimental design \cite{nguyen2024quality}, materials science \cite{liu2024diversity}, information theory \cite{nguyen2025vendi}, and algorithmic microscopy \cite{pasarkar2025vendiscope}. 
	
	In this study, we present techniques for applying the Vendi score to viral genomic data, using rich SARS-CoV-2 RNA sequence data from the United Kingdom as a paradigm. Applying the Vendi score to raw sequence data as well as phylogenetic trees and simulated data, we show how the tunability of the Vendi score (with respect to abundance weighting) allows rapid discernment of potential new viral variants, while avoiding classification-dependent artifacts present in supervised diversity measures such as \textit{Richness} and the \textit{Hill number} \cite{hill1973diversity}. While applied to SARS-CoV-2 here, the methods are fully general and may be applied to any pathogen or microbe with sufficient genomic surveillance.
	
	\section*{Results}
	\subsection*{Diversity dynamics of SARS-CoV-2}
	The sequence data obtained for SARS-CoV-2 throughout and beyond the pandemic phase is unprecedented in quantity and scope. This richness of data allowed near-real-time surveillance of the evolution of the pathogen – something that turned out to be pertinent, as the virus exhibited remarkable strain turnover \cite{duarte2022rapid,markov2023evolution}. This combination of extensive sequencing and varied evolutionary history in turn makes SARS-CoV-2 an ideal testbed for the Vendi score.
	
\begin{figure}
		\centering
		\includegraphics[width=1.0\linewidth]{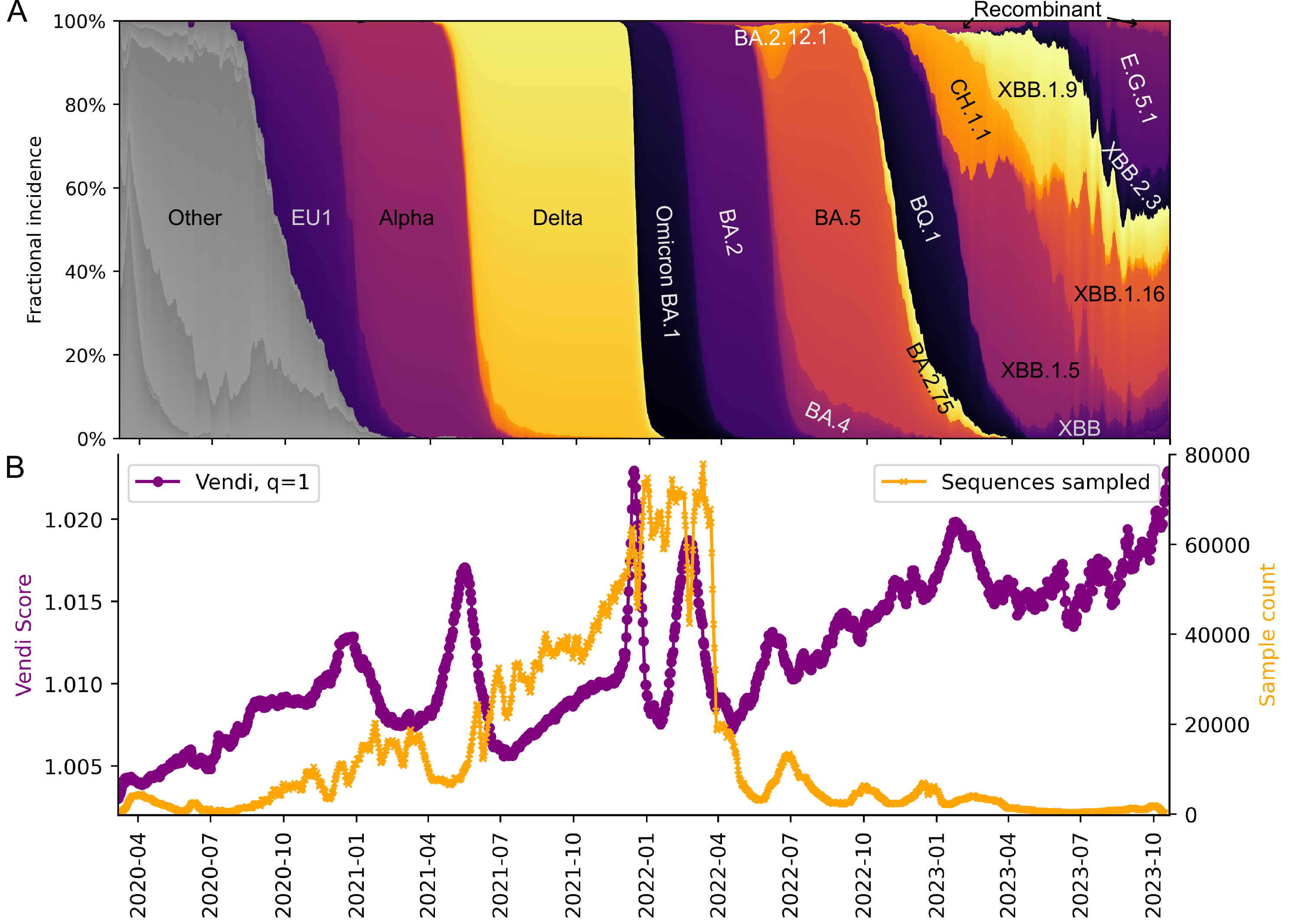}
		\caption{\textbf{SARS-CoV-2 diversity dynamics via the Vendi score.} \textbf{A)} The frequency of major SARS-CoV-2 variants in the UK through time, as a fraction of sampled sequences in each week-long window. Only variants that made up at least 1\% of sampled sequences during at least one week are included here. \textbf{B)} The Vendi score (Eq. \ref{eq:Vendiq1}) of the sampled sequences ({\color{violet}purple line}), assuming a linear similarity function $S_{ij} = 1-d_{ij}/L$ with $d_{ij}$ the number of nucleotide mismatches between sequences $i$ and $j$ and $L$ the length of the SARS-CoV-2 genome. The {\color{orange}orange line} indicates the number of sequences included in each week-long window. In computations, this was capped at 10,000 sequences. }
		\label{fig:VarTimeseries}
	\end{figure}

	In Fig. \ref{fig:VarTimeseries}, the frequencies of major SARS-CoV-2 variants (panel A) and the corresponding Vendi Score time series (panel B) are shown, based on UK sequence data made publicly available through GenBank \cite{sayers2024database,benson2013genbank}.
	As shown in Fig. \ref{fig:VarTimeseries}B, sequencing intensity has varied widely during the global health emergency phase (March 2020 \cite{Cucinotta_Vanelli_2020} to May 2023 \cite{harris2023declares}) from only a few hundred sequences per week to tens of thousands. To facilitate a direct comparison between different time points, the Vendi scores are computed on subsets of 100 sequences each, averaging across multiple such subsets when the number of available sequences in a given time window allow for it.
	
	The periods between major variant transitions are marked by gradual diversification -- see e.g. the period from July to December 2021 when the Delta variant dominated. Variant transitions themselves tend to be accompanied by a sharp increase in diversity, indicating that the emerging variant is substantially different from the resident one. This type of saltational (jump-like) evolution was observed during several of the major variant transitions of SARS-CoV-2 \cite{nielsen2023host}.
	Multiple hypotheses exist as to the origins of these jumps, with accelerated evolution associated with immunocompromised individuals currently being the most likely\cite{markov2023evolution,chen2021emergence,nielsen2023host}. It is worth noting that some significant jumps occurred before population immunity was widespread (e.g. the transition from the ancestral variant to Alpha and to a lesser extent Alpha to Delta \cite{nielsen2023host}), indicating that such jumps are not necessarily driven by selection for escape from pervasive population immunity. Such jumps are, however, not a universal feature, with later omicron sub-variants (from approx. mid-2022) not always differing strongly from their predecessors, and thus not producing pronounced diversity spikes.
	
	The sharp increase in diversity in the initial phase of a major variant transition is followed by a precipitous drop with a clear interpretation: since the emerging variant is of recent origin, its internal diversity is limited. Furthermore, a new highly fit variant tends to cause a selective sweep, pushing out other variants through competition for susceptibles, as well as due to interventions introduced in response to the new variant. Such interventions tend to bring less fit variants below an effective reproductive number of 1 before a highly fit variant is similarly affected. The effective reproductive number is the average number of new infections that each infection with a particular pathogen give rise to. If this is below 1, the prevalence of the disease in question will thus decrease. In addition to the internal diversity of an emerging variant being low due to its recent evolutionary origins, the diversity is also directly affected by the reproductive number, as explored in Supporting Fig. \ref{fig:FigS_Reff}. Consequently, the sequences belonging to a highly fit emerging variant is likely to form a low-diversity subset of the collected sequences. 
	
\begin{figure}
		\centering
		\includegraphics[width=0.85\linewidth]{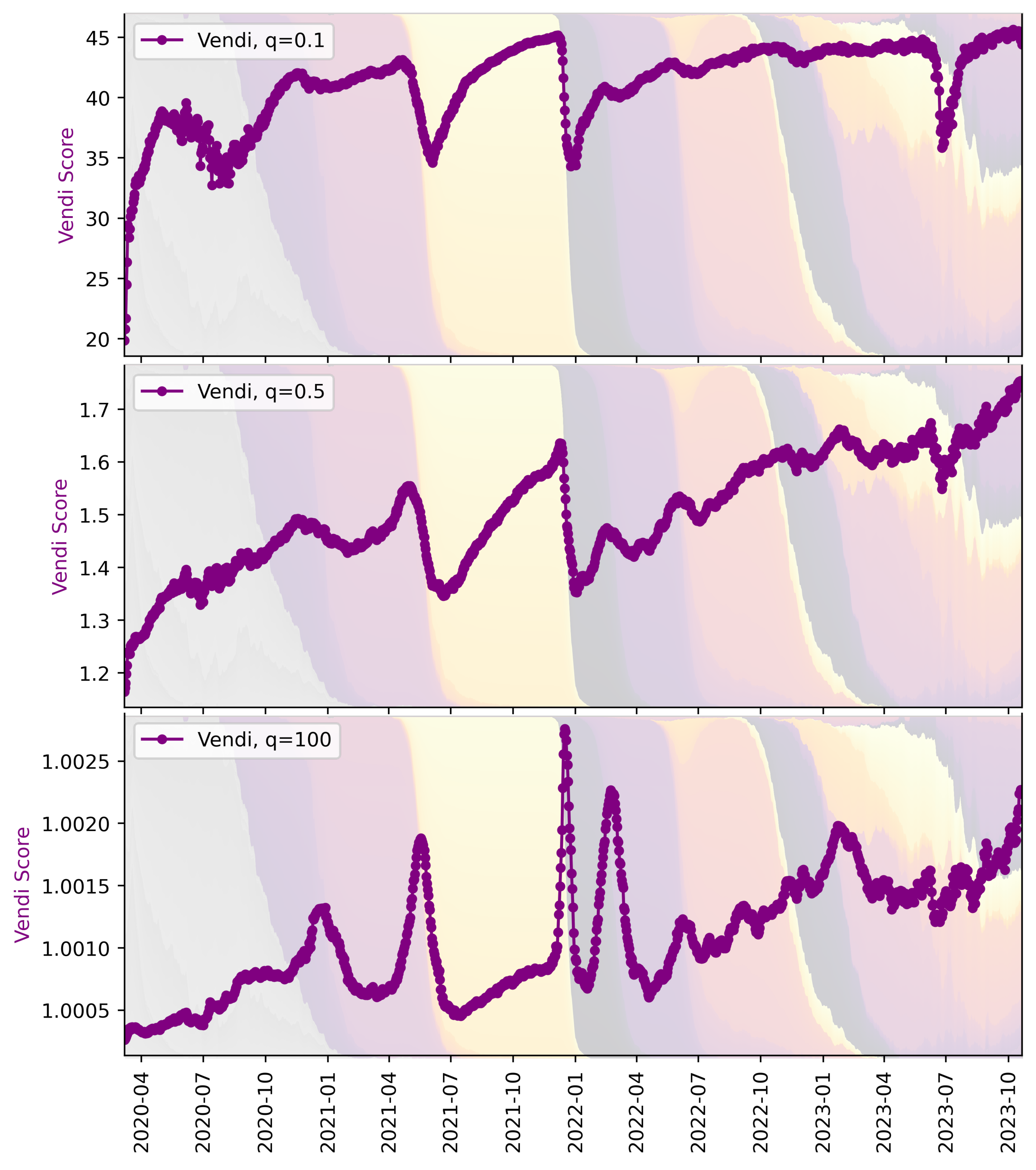}
		\caption{\textbf{At different values of $q$, the Vendi score emphasizes different aspects of SARS-CoV-2 sample composition.} At low $q$ values ($q<1$, \textbf{A-B}), more emphasis is placed on rare signals, leading to a pronounced drop in diversity when a new (initially rare) variant appears. At high $q$ values ($q>1$, \textbf{C}), variant transitions are instead marked by a spike in diversity due to the co-circulation of two (or more) distinct variants rather than a single dominant one. }
		\label{fig:Vendiq}
	\end{figure}

	The viral genomic diversity time series of Fig. \ref{fig:VarTimeseries} explores the $q=1$ Vendi score (Eq. \ref{eq:Vendiq1}). 
	However, the sensitivity parameter $q$ allows us to probe different aspects of viral diversity over time (see Eq. \ref{eq:Vendiq}). For example, a low $q$ allows for clear detection of the reduction in diversity caused by a selective sweep favouring an emerging variant. A $q$ value of $0.1$ results in a diversity time series (Fig. \ref{fig:Vendiq}A) which exhibits no sudden peaks at the emergence of a new variant. This measure thus exhibits low sensitivity to the dissimilarity between successive variants but clearly represents the low-diversity situation following a selective sweep.  
	Conversely, one may be interested in singling out the increase in diversity caused by a new variant which diverges genotypically from the previously circulating variant (Fig. \ref{fig:Vendiq}C). In e.g. influenza, antigenic distance (a determining factor in influenza strain replacement \cite{bedford2012canalization}) is known to correlate (imperfectly \cite{bedford2014integrating}) with sequence-level dissimilarity \cite{anderson2018antigenic,smith2004mapping}, spikes in which are more easily detected at large $q$.
	
	\subsubsection*{Classification independence}
\begin{figure}
		\centering
		\includegraphics[width=0.9\linewidth]{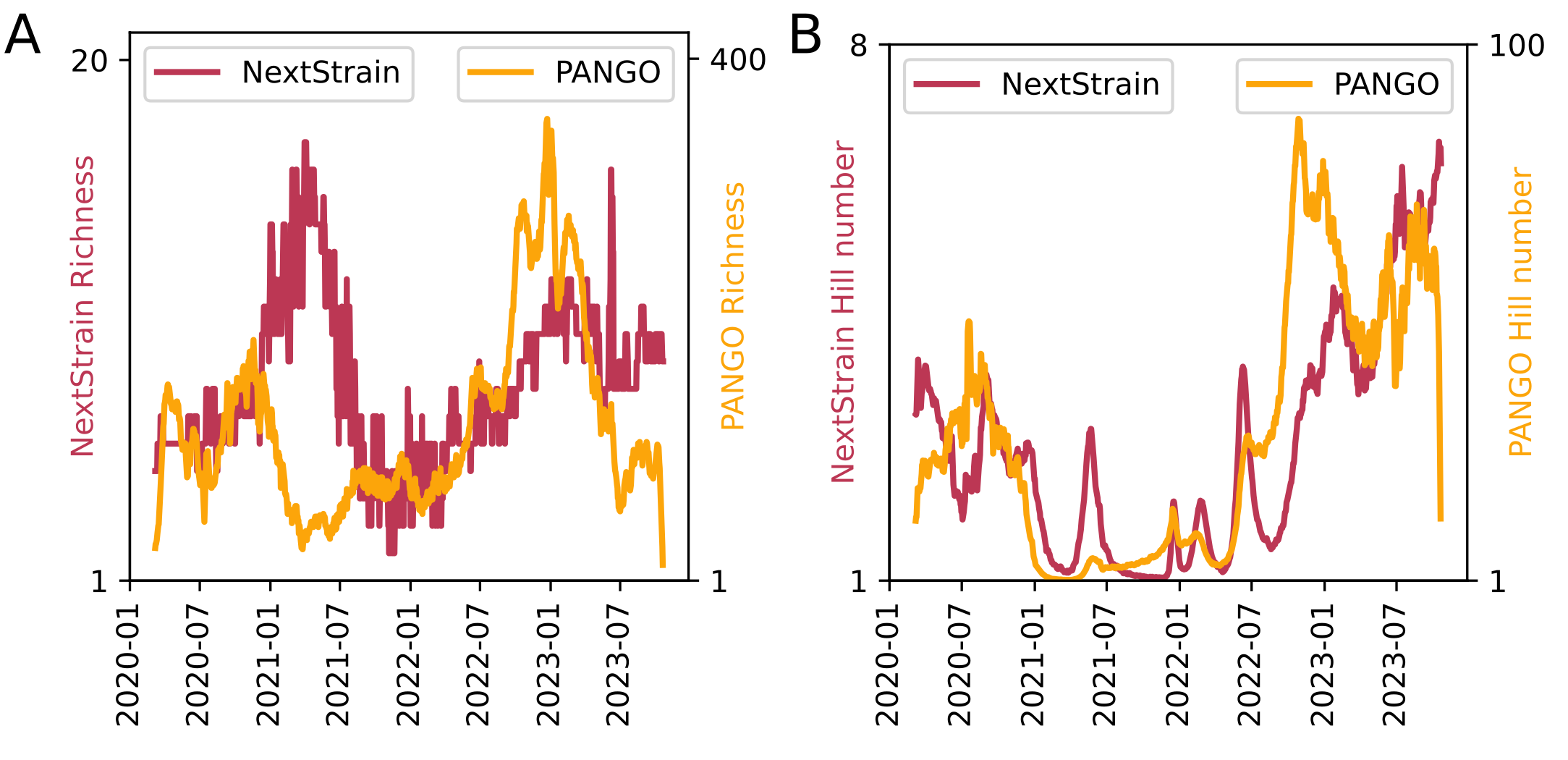}
		\caption{\textbf{Different classifications lead to significantly different diversity time series as measured by \textit{Richness} and \textit{Hill numbers}.} \textbf{A)} Richness, i.e. the number of classes represented in a given sample.
			\textbf{Red:} Richness of Nextstrain clades. \textbf{Orange:} Richness of Pango Lineages. \textbf{B) } Hill numbers, while more detailed than Richness, also yield substantially classification-dependent results. \textbf{Red:} $q=1$ Hill number using Nextstrain Clade classification \textbf{Orange:} $q=1$ Hill number using Pango Lineage classification.}
		\label{fig:RichHill}
	\end{figure}

	For SARS-CoV-2 variants, the most widely used classification system is \textit{pangolin} (Phylogenetic Assignment of Named Global Outbreak LINeages) \cite{otoole2021assignment}, with individual lineages referred to as Pango lineages \cite{otoole2022pango}. The Nextclade/\hspace{0pt}Nextstrain system \cite{aksamentov2021nextclade} is a more coarse-grained classification system widely used to designate variants and sub-variants of SARS-CoV-2. These systems have been indispensable for making sense of the multitude of SARS-CoV-2 variants, but for diversity measurements, classification comes at a cost.
	In Fig. \ref{fig:RichHill}, we explore the classification-dependence of two common diversity measures, the Richness -- the number of classes (generically: taxons/\hspace{0pt}variants/\hspace{0pt}types) present -- and the Hill number.  As Fig. \ref{fig:RichHill} shows, it matters significantly which classification is used, not only in terms of the overall diversity level, but in terms of the observed trends as well.
	While the Hill number includes abundance information and is thus more detailed than Richness, the results are still fundamentally classification-\hspace{0pt}dependent (Fig. \ref{fig:RichHill}B-D).
	
	Another facet of diversity which is not well captured by the Hill number is the internal diversification of a variant by gradual accrual of mutations -- something that can be clearly witnessed in the Vendi score, for example during the reign of the Delta variant in the latter half of 2021 (Fig. \ref{fig:VarTimeseries}B).

	\subsection*{Detecting changes in diversity: simulated data}
	In this section we apply diversity measures to idealized situations where a novel variant emerges in an already diverse background, or where several variants co-circulate at different levels of intra-variant diversity. 
	
\begin{figure}
		\centering
		\includegraphics[width=1.0\linewidth]{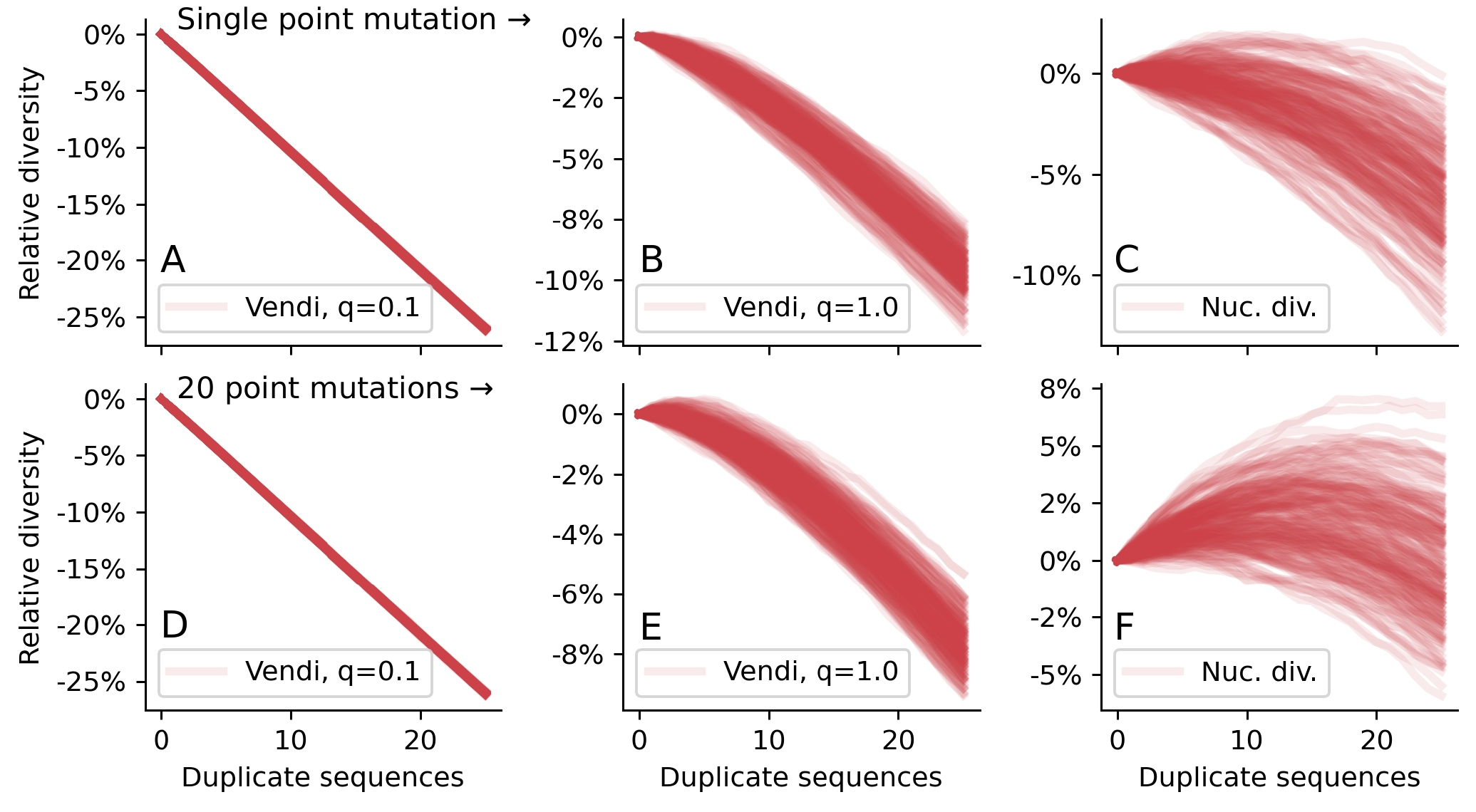}
		\caption{\textbf{The tunability of the Vendi score allows discernment of an emerging variant.} Growth of an idealized low-diversity clade is simulated by introducing duplicates of a single ``variant'' sequence in an otherwise diverse background of bit-string sequences. \textbf{A-C)} Variant arises by a single point mutation (a random bitflip is made in an existing sequence before duplicating). \textbf{D-F)} Variant arises by 20 point mutations (saltational evolution, 20 random bitflips are made in an existing sequence before duplicating).
			Constant infected population size $N=100$, genome length $L=1000$. Initially, $n\sim\text{Pois}(50)$ mutations are independently introduced in all $N$ sequences.}
		\label{fig:seqdup}
	\end{figure}

	When a new viral variant emerges in a population, it is necessarily unclassified and it is thus desirable to have measures at one's disposal diversity that \textit{1}) do not require pre-classification and \textit{2}) reflect the emergence of a new variant in a predictable manner. Existing measures which fulfill criterion \textit{1} include nucleotide diversity \cite{hartl2007principles} and mean pairwise dissimilarity (MPD) \cite{kembel2010picante,de2016functional,miller2013niche,ricotta2017beta}. In general, the MPD depends on the similarity function employed, but when using a linear similarity function, MPD is proportional to the nucleotide diversity. For this reason, we include only one of the two (nucleotide diversity) in this section.
	We explore whether the Vendi score and the nucleotide diversity satisfy criterion \textit{2} by means of a numerical simulation. In Fig. \ref{fig:seqdup}, we consider the emergence of a variant with low internal diversity in a diverse background, using the first simulation algorithm described in \nameref{sec:methods}.
	In panels A-C, the emerging variant is assumed to be closely related to already circulating viruses, having arisen by a single nucleotide change. With multiple realization of this process, it becomes clear that the nucleotide diversity may either increase or decrease as the variant proliferates, and thus does not provide a dependable method to detect the appearance of a novel variant. The $q=1$ Vendi score tends to decrease, although the pattern is initially slightly unclear. At very low $q$, however, the Vendi score decreases monotonically and thus provides a clear indication of the emergence of the new variant. 
	In Fig. \ref{fig:seqdup}D-F, the new variant is assumed to have arisen by a saltation (implemented as 20 simultaneous single point mutations). In this case, only the low-$q$ Vendi score consistently decreases.
	In Supporting Fig. \ref{fig:FigS_dup}, we repeat the analysis at while allowing for continuing accumulation of mutations, such that variant genomes are only \textit{near}-duplicates rather than perfect duplicates. The conclusion remains that the low-$q$ Vendi score is especially well-suited for detection of emerging variants.
	
\begin{figure}
		\centering
		\includegraphics[width=1.0\linewidth]{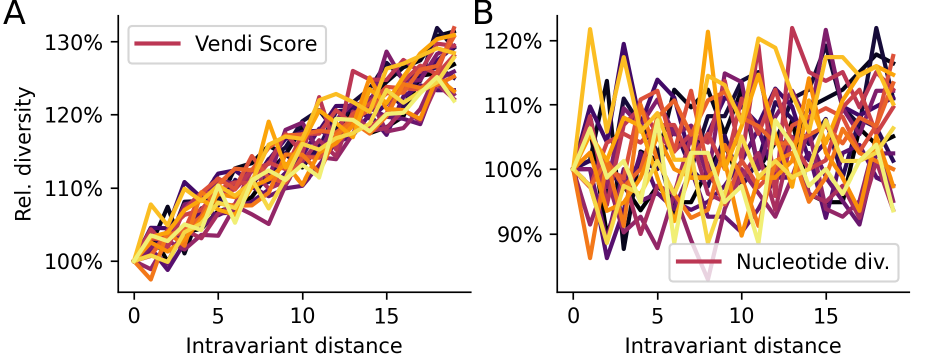}
		\caption{\textbf{The Vendi score is sensitive to within-clade diversification.} Nucleotide Diversity exhibits low sensitivity in this scenario because it does not take feature correlations into account.  Simulation in which five distinct groups of sequences (clades) diversify internally, keeping the mean genomic distance between members of \textit{different} groups constant at 50. \textbf{A)} The Vendi score ($q=1$) captures the increasing diversity in a predictable manner across simulations. \textbf{B)} Nucleotide diversity, i.e. the mean number of pairwise mismatches between all sequences.
			Bitstring genome length: $L=1000$. 
		} 
		\label{fig:intravariant}
	\end{figure}

	We now turn to the co-circulation of multiple distinct variants. Fig. \ref{fig:intravariant} maps the changes in diversity as each of five variants is made more internally diverse (by adding random point mutations) while keeping the typical genomic distances between the variants unchanged. This reveals one of the strengths of the Vendi score: it takes correlations into account \cite{friedman2022vendi} -- the entire set of $\tfrac{1}{2}n(n-1)$ internal (dis)similarities affect its value, not just the average dissimilarity. Fig. \ref{fig:intravariant} shows that the $q=1$ Vendi score captures the steady diversification, while the nucleotide diversity shows only a weak tendency and a noisy signal. Indeed, the nucleotide diversity becomes less and less sensitive to the internal diversification of each variant as the number of distinct variants increases, while the Vendi score retains sensitivity, as we show in Supporting Fig. \ref{fig:FigS_intravar}.
	Figure 2 of \cite{friedman2022vendi} explores a related concept, and shows that the Vendi score is superior to mean pairwise dissimilarity (called IntDiv in this context) in detecting per-component variance in data sampled from univariate mixture-of-normal distributions.

	\subsection*{Vendi Scoring phylogenetic trees -- novel variants as diversity-outliers}
	Until now, we have applied the Vendi score in aggregate, to the entire population under scrutiny (e.g. to all sequences collected on a given day).
	In this section, we explore the integration of the Vendi score with phylogenetic trees, allowing the evaluation of the diversity of individual clades. 
	
\begin{figure}
		\centering
		\includegraphics[width=0.9\linewidth]{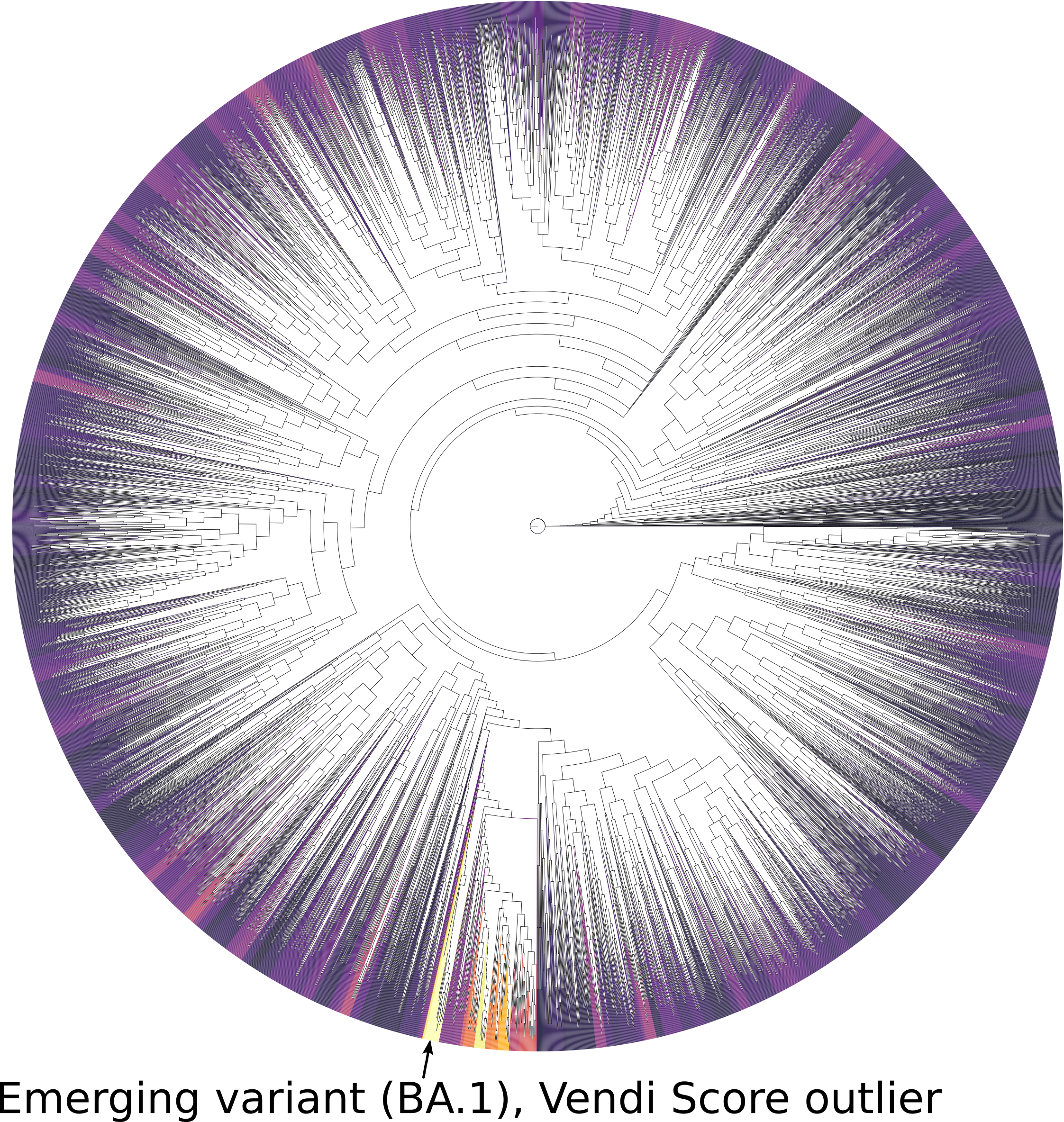}
		\caption{\textbf{Vendi scoring a phylogenetic tree reveals high and low diversity clades at a glance.} Clade-wise Vendi scored cladogram based on UK SARS-CoV-2 sequences obtained on 2021-12-05. Light yellows indicate low VS while dark purples indicate high VS. The bright yellow clade towards the bottom consist of Omicron BA.1 sequences, representing the then-invading variant. Omicron sequences make up $3.7\%$ of this data set while Delta sequences make up $96.3\%$. Visualization created with TreeViewer \cite{treeviewer}.}
		\label{fig:Tree}
	\end{figure}

	A cladewise Vendi scored phylogenetic tree is shown in Fig. \ref{fig:Tree} in the form of a cladogram. This tree is based on 6686 sequences collected on 2021-12-05 when the Omicron BA.1 variant was just emerging in the UK.
	One region of the tree appears to have much lower diversity (brighter colors) than what is typical. Upon scrutiny, the low-diversity clades turn out to correspond to the emerging Omicron BA.1 variant. Indeed, this clade has an excess diversity ($\mathit{VS}-1$) of less than half of the least-diverse Delta clade, and less than 1/27 of the most diverse clade. This Omicron clade could thus have been identified on the basis of its Vendi score absent any classification of the new variant.
	In this example, clades were scored purely on their Vendi score, as novel variants are initially expected to present as low-diversity clades. However, supplementary information could be considered for inclusion in the overall score, such as the typical distance of the clade members to the rest of the nodes, which may be relevant if new variants are also expected to be associated with significant genomic novelty. 
	In Supporting Fig. \ref{fig:FigS_tree}, a Vendi-scored phylogeny from Nov. 5, 2020 is included which singles out the then-emerging Alpha variant as the lowest-diversity clade before it had been formally classified. 
	These examples showcase how the Vendi score may serve as an adjunct tool in viral variant surveillance, supplementing traditional epidemiological methods.

	\section*{Limitations and practical considerations}
	While it is true that the SARS-CoV-2 pandemic presented unprecedented availability of sequence data, there are some limitations of diversity measures such as the Vendi score that are not overcome by sheer data quantity. Sampling heterogeneity is an example of such a limitation -- if samples are collected in an uneven manner, such that clusters of closely related sequences are overrepresented in the data, these will appear as low-diversity groups of sequences without any real evolutionary significance.
	In general, representativeness of samples is a challenge, especially when national and regional sequencing efforts vary widely \cite{brito2022global,berrig2022heterogeneity}. 
	As we have shown, the Vendi score is highly flexible and allows probing different aspects of diversity by tuning $q$, and by choosing a suitable similarity kernel. This flexibility, however, means that appropriate choices must be made for each data set.
	
	\section*{Discussion}
	In connection with the 2014-2016 Ebola outbreak in West Africa, Quick et al. \cite{quick2016real} noted that \textit{``Sequence data may be used to guide control measures, but only if the results are generated quickly enough to inform interventions.''}
	In that spirit, we propose that unsupervised, ready-to-use sequence-based metrics such as the viral Vendi score can play an important role in timely surveillance of pathogens. By design, the Vendi score requires only the genomic data itself (along with a suitable similarity function), making it well suited to real-time, large-scale analyses that can complement existing frameworks. \\
	Although some early studies characterized SARS-CoV-2 as displaying ''minimal diversity`` \cite{dearlove2020sars}, and even speculated that low diversity was an Achilles heel of the virus \cite{rausch2020low}, it soon became clear that SARS-CoV-2 had a remarkable capacity for generating new variants. 
	Multiple tools were created to classify and track evolving SARS-CoV-2 lineages, chief among which are the Pango, Nextstrain and WHO VOC/VOI/VUM (variant of concern/of interest/under monitoring) classifications \cite{rambaut2020dynamic,otoole2022pango,focosi2023sars,world2023updated}. These systems have criteria for lineage designation that largely focus on epidemiological significance -- e.g. circulation frequency (regionally/globally) and growth rates. The initial proposal for the Pango nomenclature \cite{rambaut2020dynamic} lays out a set of criteria for designating a new lineage, the upshot of which is that a potential new lineage must be associated with spread into a geographically distinct population (relative to its ancestor), have a minimum of one defining nucleotide change and exhibit a certain phylogenetic likelihood.
	Once a new sequence is added, the machine learning-based pangoLEARN software may be used to determine the lineage into which it belongs \cite{otoole2021assignment}. Nextstrain's criteria also stress epidemiological significance, by requiring e.g. at least two months of circulation at a frequency of $>20\%$.\cite{focosi2023sars} While such criteria, which rely in part on monitoring spread globally and regionally, are utterly sensible and the Pango and Nextstrain systems have been resounding successes, surveillance may benefit from complementary diversity measures which are unsupervised and 
	sensitive to within-lineage variability.
	When deployed in tandem with existing classifications, this approach may yield early signals of emerging variants, flagged by changes in diversity that standard lineage criteria might not immediately capture.
	
	Among the principal strengths of the Vendi score are its flexibility and computational efficiency. Here, we used a linear kernel to define sequence similarity (via simple Hamming/\hspace{0pt}Levenshtein distances), for reasons of parsimony. 
	However, not all mutations are created equal, and need not be weighted equally in the calculation of the Vendi score. For example, nucleotide similarity matrices often assign transitions (purine$\leftrightarrow$purine and pyrimidine$\leftrightarrow$pyrimidine substitutions) higher similarity than transversions (purine$\leftrightarrow$pyrimidine) \cite{xia2018sequence}, and substitutions are often assigned higher similarity than are insertions and deletions. At the amino acid level, scoring matrices frequently assign individual similarities to each possible amino acid pair, e.g. BLOSUM62 \cite{eddy2004blosum62}. 
	In a similar and perhaps even more salient vein, information about the antigenic -- or, more generally, phenotypic -- significance of changes at individual sites may be included in the Vendi score to account for changes in, for example, immune response or binding (using e.g. deep mutational scanning \cite{starr2022shifting}). The result would thus be a specialized functional diversity measure \cite{dieng2025unified}. An antigenically-\hspace{0pt}informed Vendi score may thus allow flagging not only genomic novelty, but also immune-escape potential.
	
	Although this study focuses on SARS-CoV-2, diversity measurements are equally central to numerous other pathogens. For example, it is a long-standing puzzle that influenza A (H3N2), a pathogen undergoing rapid evolution, exhibits very limited standing diversity (antigenically and genotypically) at any given time, despite experiencing strong pressure to evolve ``away'' from human immunity \cite{bedford2012canalization,koelle2006epochal}. Studies of genomic diversity are also highly useful for understanding the evolutionary history of different influenza types, as well as their circulation patterns in different hosts \cite{nielsen2024one}.
	At the within-host level, viral diversity also plays an important role. For example, it has been found that intrahost nucleotide diversity of human respiratory syncytial virus (RSV) varies between antigenic subtypes (RSV A and B) \cite{lin2021distinct}, and that diversity is correlated with immune pressure \cite{grad2014within}. Revisiting data sets such as these using a measure that takes feature correlations into account, rather than just the mean number of nucleotide mismatches, appears promising. 
	
	Beyond viruses, studies of microbial communities, such as the vertebrate gut microbiome, stand to benefit from classification-agnostic diversity measures like the Vendi score. In recent decades, the characterization of the diversity of microbial communities has taken on increasing importance \cite{gibbons2015microbial,pace1997molecular,hunter1998value}. 
	Gut microbial diversity in particular is known to be associated with disease \cite{durack2019gut}, host fitness in general \cite{pfau2023social}, cognitive and behavioral outcomes\cite{sarkar2020role,pfau2023social,johnson2020gut,canipe2021diversity} and has changed rapidly during human evolution \cite{moeller2014rapid,moeller2017shrinking}. Studies of the vertebrate gut microbiome have generally employed within-population diversity (\textit{Alpha diversity}) measures based on either the number of taxa, or the abundances of each taxon, but given the richness of microbiome genomic data, detailed (intraspecific) diversity measurements are a promising area of study \cite{sanders2022low}.

	\section*{Conclusion}
	With pathogen genomic data collection during disease outbreaks now occurring at an
	unprecedented scale and speed, tools which allow for rapid analysis are paramount. In this work, we have provided techniques which allow the Vendi score to be applied, in real-time, to incoming genomic sequences for a given pathogen, requiring nothing beyond the sequence data itself and a suitable similarity function. The Vendi score shows promise as a supplementary tool for detection of emerging viral variants, both through time-series based analysis of diversity dynamics and via integration with phylogenetics.
	
	\section*{Materials and methods}\label{sec:methods}
	\subsection*{The Vendi score}
	Here we provide a brief summary of the Vendi score as originally developed in \cite{friedman2022vendi}, further mathematical properties can be found in that paper. Given a collection $\{X_i\}_{i=1,\dots,n}$ of samples and a positive semi-definite similarity kernel function $K_{ij} = K(X_i,X_j)$, the Vendi score of the collection is given by:
	\begin{align}
		\mathit{VS}_1 &= \exp\left(-\sum_{i=1}^n \lambda_i \log(\lambda_i) \right), \label{eq:Vendiq1}
	\end{align}
	where the $\lambda_i$ are the eigenvalues of the normalized similarity matrix  $\boldsymbol{K}/n$. The choice of similarity function is discussed in the next section.
	
	The subscript $1$ above ($\mathit{VS}_1$) indicates that this is a special case of a larger family of Vendi scores \cite{pasarkar2023cousins}, having a tunable parameter $q$:
	\begin{align}
		\mathit{VS}_q &=\left(\sum_{i=1}^n \lambda_i^q\right)^{1/(1-q)}. \label{eq:Vendiq}
	\end{align}
	The parameter $q \geq 0$ allows emphasis to be placed on rarer or more abundant types in the data set. While $q=1$ cannot be directly substituted in \eqref{eq:Vendiq}, it does hold that $\lim_{q\to 1} \mathit{VS}_q = \mathit{VS}_1$.
	
	The Vendi score may be compared with the Hill number \cite{hill1973diversity}, often referred to as the \textit{true diversity} within ecology. Given a set of \textit{species} $i \in \{1,\dots,R\}$ with relative abundances $p_i$, the Hill number of order $q$ is defined by
	\begin{align}
		^qD &= \left(\sum_{i=1}^R p_i p_i^{q-1} \right)^{1/(1-q)},
	\end{align}
	which may also be recognized as $M^{-1}_{q-1}$, i.e. the reciprocal of the weighted generalized mean abundance of order $q-1$.
	In the $q=1$ case, where species are weighted proportional to their abundance (favouring neither rare nor abundant species), the Hill number reduces to the exponential of the Shannon entropy:
	\begin{align}
		^1D &= \exp\left(-\sum_{i=1}^R p_i \log(p_i) \right).
	\end{align}
	Similarly, the $q=1$ Vendi score (Eq \ref{eq:Vendiq1}) may be recognized as the exponential of the von Neumann entropy from quantum statistical mechanics, of a density matrix $\boldsymbol{\rho} = \boldsymbol{K}/n$. As with the Hill number \cite{hill1973diversity}, the value of the sensitivity parameter $q$ controls the weight given to rare and abundant types in the Vendi score \cite{pasarkar2023cousins}. At large $q$, the most abundant types dominate and the Vendi score tends towards $1/\lambda_\mathrm{max}$ with $\lambda_\mathrm{max}$ being the dominant eigenvalue of the reduced similarity matrix $\textbf{K}/n$. At infinite $q$, there are thus effectively only two types: the most abundant one, and everything else.
	At low $q$, the opposite situation arises. As $q$ is decreased, the weighting of different types becomes more and more similar. As $q$ tends to zero, the Vendi score thus tends to an integer $m \leq n$ which counts the number of dissimilar types, however minute the differences between them.
	
	In this study, we will often compare the Vendi score with other diversity measures, chiefly the \textbf{Hill number} (defined above), the \textbf{Mean Pairwise Dissimilarity}, MPD (as defined below), 	the \textbf{nucleotide diversity} (mean distance between genomes in a set) and the \textbf{Richness} $R$, defined as the number of types (i.e., species) present in a set.
	The MPD (also known as IntDiv \cite{friedman2022vendi,hu2024hamiltonian}, when applied to molecules) is given by 
	\begin{align}
		\mathit{MPD} = 1-\frac{1}{n^2}\sum_i \sum_j K_{ij}.
	\end{align}
	
	\subsubsection*{The Vendi score as an effective number}
	Some diversity measures (such as mean pairwise dissimilarity, MPD) are constrained to a fixed interval (e.g. $[0,1]$) while others (such as nucleotide diversity) have no universal maximum value (independent of sequence length) for a sample set of size $n$. The Vendi score, however, attains its maximal value of $n$ when all eigenvalues $\lambda_i$ are equal in magnitude, $\lambda_i=1/n$. At the opposite extreme, the lowest possible Vendi score of $1$ is attained when only a single eigenvalue is non-zero.  The Vendi score is thus not just a diversity measure, but belongs to a distinguished class known as \textit{effective numbers}~\cite{leinster2012measuring}. The best known effective number (of species) is perhaps the Hill number. For the Hill number, the expression of being an effective number takes the following form: if the Hill number of a system is $h$, that system is as diverse as one made up of $h$ equally abundant species. For the Vendi score, the statement must be modified to take similarity into account: if the Vendi score of a system is $v$, that system is as diverse as one made up of $v$ completely dissimilar samples.
	
	\subsection*{Application to viral sequence data}
	\subsubsection*{The similarity function}
	In order to apply the Vendi score to viral RNA sequences, a notion of similarity must be defined. 
	Since the present work is concerned with viral evolutionary epidemiology, genomic distance in the form of the number of nucleotide mismatches between sequences, serves as a natural starting point for defining a similarity measure. We will primarily use a linear similarity metric, $K_{ij}=1-d_{ij}/L$, where $L$ is the genome length (measured in base pairs) and $d_{ij}$ is the simple (unweighted) Hamming\footnote{Note that, since we are including nucleotide deletions, insertions and substitutions in our calculations of the Hamming distance $d_{ij}$ (equivalent to including alignment gap characters in the counting), $d_{ij}$ is equal to the Levenshtein distance, and the two terms may thus be used interchangeably in this context.} distance between genomes -- i.e. the number of nucleotide mismatches. However, we note that it is entirely possible to define a measure which weighs individual mismatches according to e.g. their phenotypic importance. Epidemiologically pertinent phenotype-associated information could originate from antigenic cartography, receptor binding studies (\textit{in silico} or otherwise) or simply the position of the site in question relative to known epitopes, resulting in a functional diversity measure \cite{dieng2025unified}.
	
	In addition to the linear similarity measure used in this paper, other positive-semidefinite measures may be considered, for example the exponential $K_{ij} = \exp\left(-d_{ij}/\sigma\right)$, with $\sigma$ a free parameter \cite{hutter2014algorithm}. This measure has the advantage of allowing a tunable, nonlinear dependence on genomic distance. On the other hand, the absence of any free parameters in the linear similarity measure makes for unambiguous interpretability. 
	
	\subsubsection*{Vendi scoring SARS-CoV-2 sequences}
	As  described above, the computation of the Vendi score is straightforward once a similarity matrix $\boldsymbol{K}$ has been computed. Here we detail the process for constructing a time series of Vendi scores of weekly (or daily) SARS-CoV-2 sequences. We base this description on the \textit{open} data sets of SARS-CoV-2 sequences offered by Nextstrain in collaboration with GenBank \cite{hadfield2018nextstrain,benson2013genbank}. The process involves the following steps:
	\begin{enumerate}
		\item Identify sequences for each time window of interest (we will use a one-week moving window at a time resolution of one day, for definiteness)
		\item Compute all pairwise genomic distances within each time window
		\item Compute a similarity matrix for each time window
		\item Lastly, calculate a Vendi score based on each of the similarity matrices
	\end{enumerate}
	A metadata filtering tool (\texttt{metadata\_extractor.py}) is provided to select the sequences of interest from the open data set. Computations are made more efficient by the fact that the full sequences are not needed, but only the changes relative to a reference sequence, since these are sufficient for the computation of the genomic distance between any pair. \texttt{metadata\_extractor.py} thus saves information on substitutions, deletions and insertions rather than the full sequences.
	A stand-alone program written in C++, \texttt{dmat}, is provided to compute all pairwise distances for a given time window, as well as a script to parallelize this task. Lastly, the Vendi score time series may be computed using \texttt{VScore\_pandemic.py}.
	
	\subsubsection*{Vendi on phylogenetic trees}
	In this section we introduce a method for Vendi scoring a phylogenetic tree -- that is, the the assignment of a diversity score to each clade of an existing phylogenetic tree. The computation of the pylogenetic tree itself is independent of the Vendi score and relies on well-known methods given below. The only real requirement is that the generated tree is output in Newick tree format.  It must be emphasized that diversity evaluation on trees requires the inclusion of either the entire set of sequences for a given locale and period of interest, or a strictly representative subsample. In particular, common practices such as removal of (near-)duplicates will skew diversity measurements and must be avoided.
	
	Our pipeline uses IQ-Tree 2 for maximum-likelihood phylogeny inference \cite{minh2020iq}, the NCBI Datasets command-line interface \cite{oleary2024exploring} for fetching any sequences not found in the Nextstrain open sequence sets \cite{hadfield2018nextstrain}, SeqKit for sequence filtering \cite{shen2024seqkit2}, MAFFT for sequence alignment \cite{katoh2013mafft}, and zstd for data compression \cite{collet2021rfc}.
	
	Clade-wise computation of the Vendi score is performed using the \texttt{VendiTree} tool, which takes as inputs a tree in Newick format as well as a pre-computed matrix of pairwise genomic distances between samples. The output is a list of clades (defined by their member sequences) and associated Vendi scores. Since the Vendi score, or indeed any diversity measure, is only really meaningful when the population (or collection of samples) in question has several members, a minimum clade size for Vendi computation must be set (default value: 20 sequences).
	
	\subsection*{Simulations -- synthetic data}
	Simulated data allow the showcasing and benchmarking of the Vendi score in a controlled environment. In this section, we detail the simulations used in Figs. \ref{fig:seqdup} and \ref{fig:intravariant} to probe the sensitivity of the Vendi score to the presence of low-diversity clades, and to intra-variant diversity, respectively.
	\subsubsection*{Simulation: low-diversity sequence subset}
	Here, we detail the simulation algorithm behind Fig. \ref{fig:seqdup}.\\
	The algorithm operates on a collection of bitstring sequences $\boldsymbol{X} = \{X_i\}_{i \in 1,\dots,N}$, each of length $L$. We denote the value of the $k$'th site of the $i$'th sequence by $X_i[k]$. Initially, set all sequences in $\boldsymbol{X}$ equal to an (arbitrary) sequence $X_0$. Without loss of generality, $X_0$ may be chosen as the all-zero sequence. An initial background level of diversity is then introduced by independently bit-flipping sites by letting $X_i[k] \to 1-X_i[k]$ with probability $p_\text{mut}$, for each $i \in \{1,\dots,N\}$ and $k \in \{1,\dots,L\}$.
	
	An increasing number of duplicate sequences are then introduced as follows, to simulate the growth of a low-diversity clade. First, choose one parent sequence $X_p$. Then introduce a number $m$ of ``variant-defining'' bit-flip mutations in $X_p$ using the method described above ($m=1$ for a single defining point mutation, e.g. $m=20$ for a saltation). 
	Let $n=1$ and iterate the following until the desired maximal number of duplicates $c_\text{max}$ is reached.
	\begin{enumerate}
		\item If $n=1$, let $X_n=X_p$. Otherwise, let $X_n=X_m$ for $m<n$.
		\item Compute all pairwise distances $d_{ij} = \text{Hamming}(X_i, X_j)$ and the corresponding similarity matrix $\boldsymbol{K}$.
		\item Compute and store diversity scores: $\mathit{VS}_q(\boldsymbol{X})$ and $\text{NucDiv}(\boldsymbol{X})$
		\item Let $n \to n+1$
		\item If number of copies $< c_\text{max}$, go to step 1.
	\end{enumerate}
	In the supplement (Supporting Fig. \ref{fig:FigS_dup}), we consider the situation where sequences undergo continual mutation, meaning that a random (binomial) number of mutations are introduced in each genome for each iteration of the above loop (between steps 1 and 2, for example).
	\subsubsection*{Simulation: intravariant distance}
	Here, we detail the simulation algorithm behind Fig. \ref{fig:intravariant}.
	As in the above case, the algorithm operates on a collection of bitstring sequences $\boldsymbol{X} = \{X_i\}_{i \in 1,\dots,N}$, each of length $L$. However, the sequences are now also members of $N_p$ distinct populations (``variants''). Denote the desired baseline inter-variant distance by $d_\text{inter}$ and the \textit{intra}-variant distance by $d_\text{intra}$. For each $d_\text{intra}$ value of interest, perform the following steps:
	\begin{enumerate}
		\item Set all sequences in $\boldsymbol{X}$ equal to an (arbitrary) sequence $X_0$. Without loss of generality, $X_0$ may be chosen as the all-zero sequence.
		\item \textbf{Inter-variant distance:} For each population, pick one member $X_p$ and perform independent bit-flips on $X_p$ with probability $d_\text{inter}/(2L)$ per site. Set all members of the populaton equal to $X_p$
		\item \textbf{Intra-variant distance:} For each sequence $X_i \in \boldsymbol{X}$, perform independent bit-flips on $X_i$ with probability $d_\text{intra}/(2L)$ per site.
	\end{enumerate}
	Note that the above algorithm only precisely produces populations with a mean intra-variant distance $d_\text{intra}$ and a mean distance between members of different populations of $d_\text{intra} + d_\text{inter}$ when both $d_\text{inter}\ll L$ and $d_\text{intra}\ll L$. To generate Fig. \ref{fig:intravariant}, $d_\text{intra}$ was varied from 0 to 20 while keeping $d_\text{intra} + d_\text{inter}=50$.
	
	\section*{Supporting information}

	\section*{Acknowledgments}
	BFN acknowledges financial support from the Carlsberg Foundation (grant no. CF23-0173). BTG would like to acknowledge support from Princeton Catalysis and Princeton Precision Health. ABD acknowledges support from Princeton Precision Health. APP is supported by an NSF Graduate Research Fellowship. 
	
	\section*{Code availability}
	All relevant code is accessible at the GitHub repository \url{https://github.com/BjarkeFN/ViralVendi}.
	
	\providecommand{\noopsort}[1]{}
	
	\newpage
	
	\section*{SI Appendix}
	\label{S1_Appendix}
	\renewcommand\thefigure{S\arabic{figure}}  
	\renewcommand\thetable{S\arabic{table}}  
	\renewcommand\theequation{S\arabic{equation}}  
	\setcounter{figure}{0}
	\setcounter{table}{0}
	
	\begin{figure}[h!]
		\centering
		\includegraphics[width=0.6\linewidth]{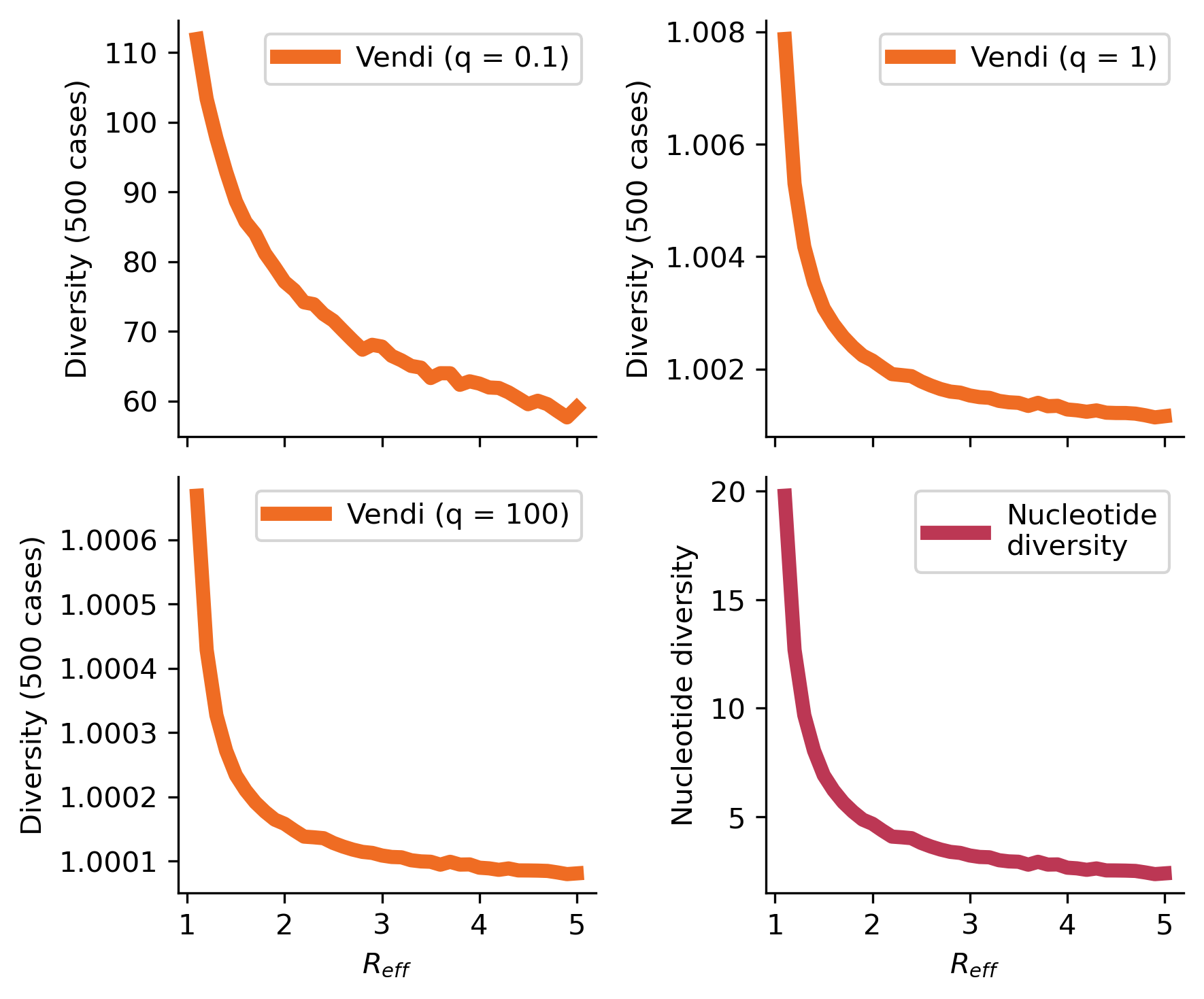}
		\caption{Under neutral evolution, observed diversity depends strongly on the reproductive number. For each value of $R_\text{eff}$, 100 branching processes with mean offspring number equal to $R_\text{eff}$ are simulated, each starting from a single individual. Each new infection was associated with a bitstring genome of length $L=500$, which is inherited at transmission. At each transmission, a random mutation (bitflip) is introduced with probability $\mu=0.3$. Simulations are run until a generation size of $500$ is reached. The genomic diversity of this last generation is then computed. The plot shows the mean of 100 simulations for each value of $R_\text{eff}$.}
		\label{fig:FigS_Reff}
	\end{figure}
	
	\begin{figure}[h!]
		\centering
		\includegraphics[width=1.0\linewidth]{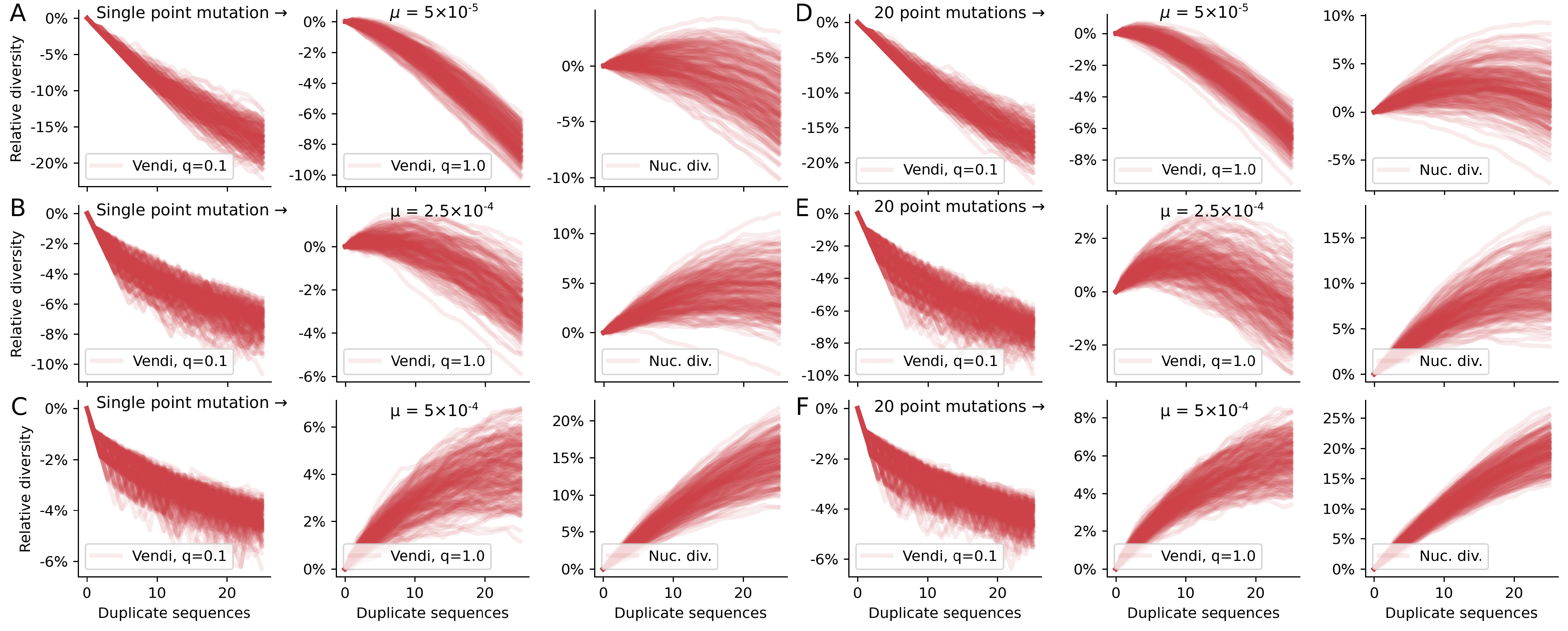}
		\caption{Supporting figure related to main text Figure 4. Growth of an idealized low-diversity clade is simulated by introducing \textbf{near-}duplicates of a single ``variant'' sequence in an otherwise diverse background of bit-string sequences. \textbf{A-C)} Variant arises by a single point mutation (a random bitflip is made in an existing sequence before duplicating). \textbf{D-F)} Variant arises by 20 point mutations (saltational evolution, 20 random bitflips are made in an existing sequence before duplicating). \textbf{In contrast to Figure 4}, further mutations are introduced in all genomes at the time of duplication (with a probability $\mu$ per site per duplication, with values indicated in each panel). This means that duplication is imperfect, emulating continuing diversification during the growth of the low-diversity clade.  
			Constant infected population size $N=100$, genome length $L=1000$. A background level of diversity is ensured by initially introducing $n\sim\text{Pois}(50)$ mutations independently in all $N$ sequences. }
		\label{fig:FigS_dup}
	\end{figure}
	
	\begin{figure}[b]
		\centering
		\includegraphics[width=0.6\linewidth]{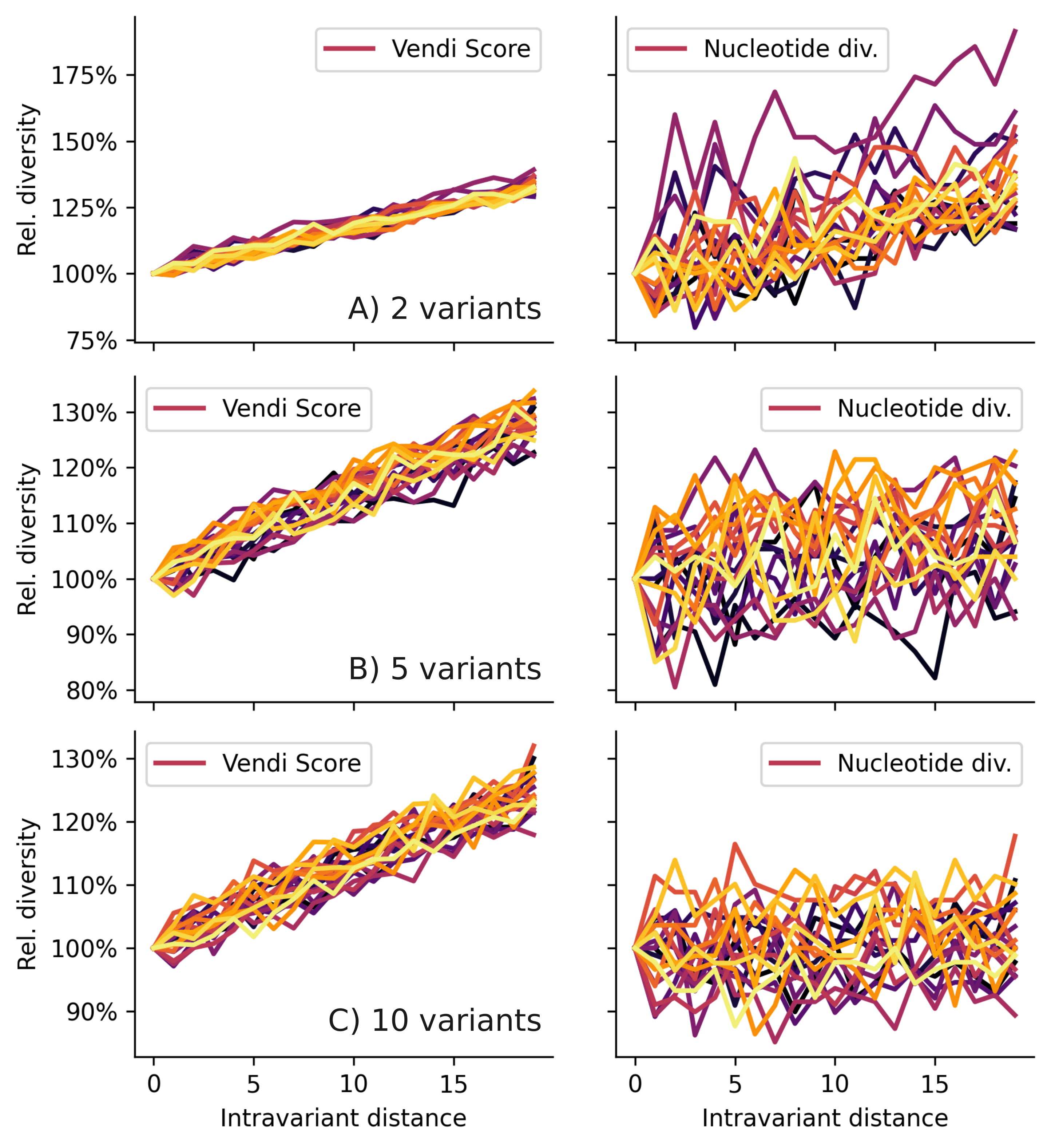}
		\caption{Supporting figure related to main text Figure 5. The Vendi score retains sensitivity to intravariant distance as the number of variants increases, while the nucleotide diversity does not.}
		\label{fig:FigS_intravar}
	\end{figure}
	
	\begin{figure}[b]
		\centering
		\includegraphics[width=0.7\linewidth]{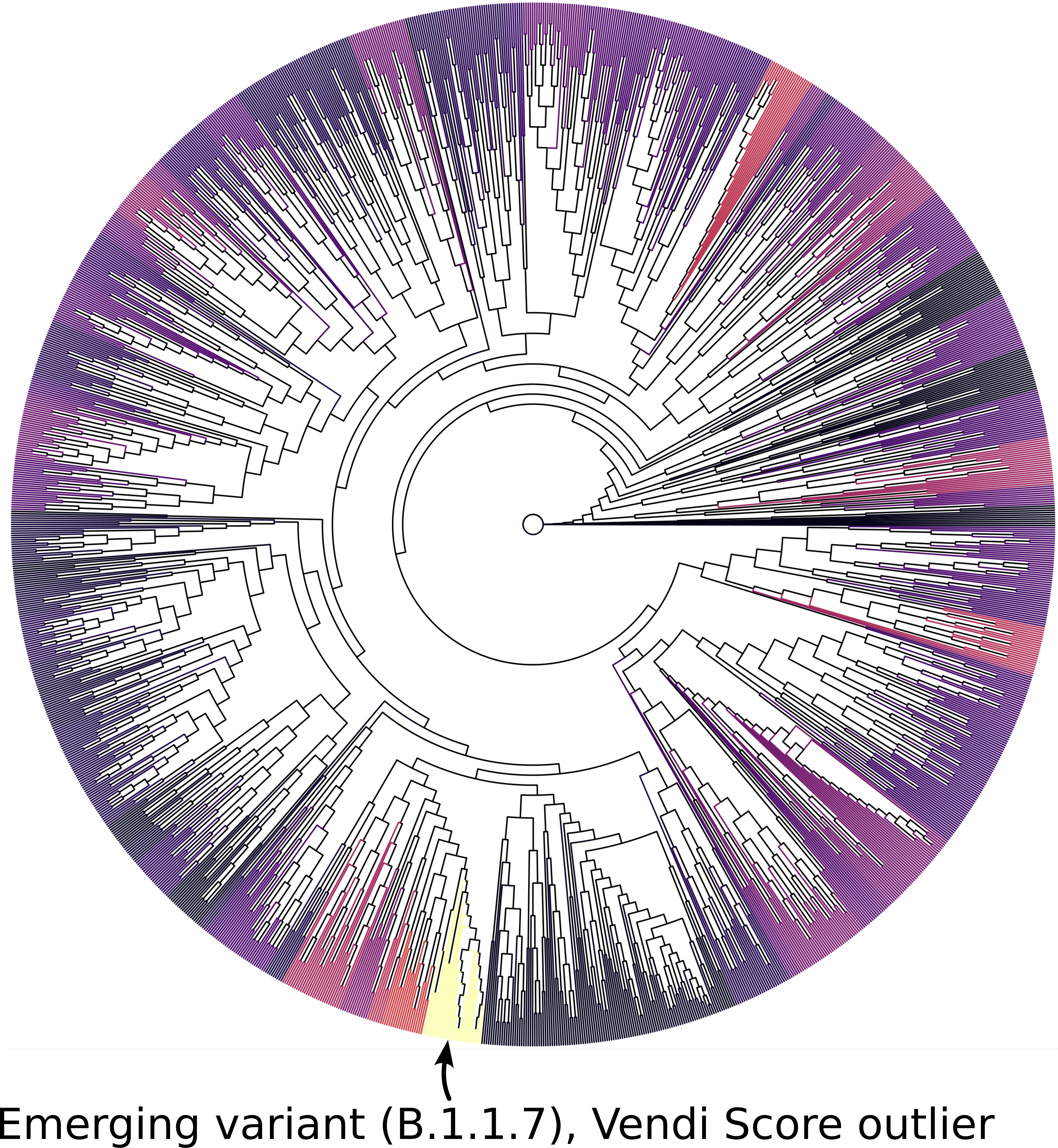}
		\caption{Supporting figure to main figure 6. Clade-wise Vendi scored cladogram based on 1485 UK SARS-CoV-2 sequences obtained on 2020-11-05. Light yellows indicate low VS while dark purples indicate high VS. The bright yellow clade towards the bottom consist of B.1.1.7 (Alpha) sequences, representing the then-invading variant. B.1.1.7 sequences make up $6.5\%$ of this data set. Visualization created with TreeViewer \cite{treeviewer}.}
		\label{fig:FigS_tree}
	\end{figure}

\end{document}